# EFFICIENT SUPPORT COUPLED FREQUENT PATTERN MINING OVER PROGRESSIVE DATABASES

Keshavamurthy B.N. , Mitesh Sharma and Durga Toshniwal

Department of Electronics and Computer Engineering,
Indian Institute of Technology,
Roorkee, Uttarkhand, India.

bnkeshav123@gmail.com, mitusuec@iitr.ernet.in, durgafec@iitr.ernet.in

## ABSTRACT

There have been many recent studies on sequential pattern mining. The sequential pattern mining on progressive databases is relatively very new, in which we progressively discover the sequential patterns in period of interest. Period of interest is a sliding window continuously advancing as the time goes by. As the focus of sliding window changes , the new items are added to the dataset of interest and obsolete items are removed from it and become up to date. In general, the existing proposals do not fully explore the real world scenario, such as items associated with support in data stream applications such as market basket analysis. Thus mining important knowledge from supported frequent items becomes a non trivial research issue. Our proposed novel approach efficiently mines frequent sequential pattern coupled with support using progressive mining tree.

## 1. INTRODUCTION

In recent years, due to the advancement of computing and storage technology, digital data can be easily collected. It is very difficult to analyze the entire data manually. Thus a lot of work is going on for mining and analyzing such data using data mining techniques. This processing of analyzing data from different techniques has a great application in the business world. For example, knowledge mined from data can be used to make strategic business decisions which help in increasing revenue, cutting cost etc. Data mining can be defined as the process of "mining" knowledge from large amount of data [1][2].

Of the various techniques of data mining analysis, sequential data/pattern analysis is one of the active areas of research work. Sequential pattern mining was first introduced in [1] as: "Given a sequence database, where each sequence consists of a list of ordered item sets containing a set of different items, and a user defined minimum support threshold, sequential pattern mining is to find all sub-sequences whose occurrence frequencies are no less than the threshold from the set of sequences."

**Definition 1.** Let $X =\{x1, x2, x3… xn\}$ be a set of different items. An element $e$, denoted by $< x1, x2,. . .>$, is a subset of items belonging to $X$ which appear at the same time. A sequence $s$, denoted by $< e1 ; e2 ; . . . ; em >$ , is an ordered list of elements. A sequence database $DB$ contains a set of sequences, and $| DB |$ represents the number of sequences. in $DB$. A sequence $α$

= < *a1 ; a2 ; . . . ; an* > is a subsequence of another sequence *β* =< *b1 ; b2 ; . . . ; bm* > if there exist a set of integers, $1 \leq i1 \leq i2 \leq in \leq m$, such that *a1* is a subset of *bi1*; *a2* is a subset of *bi2* ; .. . and *an* is a subset of *bin* [3].

Sequential pattern mining can be applied to static databases, where data do not change over time. While in many domains, the content of the databases are updated incrementally. In order to get all the sequential patterns, the mining algorithms has to run whenever database changes because some data sequences which are not frequent in old data bases may become an frequent in updated database. Thus sequential pattern mining with the incremental database corresponds to the mining process where there are new data arriving over database.

Both mining frequent items on sequential and incremental data sets have been studied extensively. However, progressive sequential pattern mining which discover sequential patterns is a new area of research. Progressive databases have posed new challenges because of the following inherent characteristics such as it should not only add new items to the existing database but also removes the obsolete items from the database.

It is thus a great interest to mine items that are currently frequent in progressive databases but coupling of support to items proposed to discover more important knowledge which plays a significant role in real world application. For example, the market basket analysis of customer who visits the supermarket may not always buys single item which are of in general day to day usage, such as egg, butter and bread .

In analysis of real world applications such as retail-shop coupling of support to an item discover more knowledge. Our contribution of this paper is to extract the supported frequent patters from progressive sequential databases.

The reminder of this paper is organized as follows: section 2 gives a formal definition of the problem definition of this paper and discusses the literature review. In section 3, explores the operation on support progressive sequential. In section 4, includes performance evaluation. We conclude our work in section 5.

## 2. PRELIMINARIES

### 2.1 PROBLEM STATEMENT

There are many research papers which have discussed the progressive databases but the existing proposals do not extract the important hidden knowledge such as supported items over progressive databases. The proposed work of this paper addresses this issue very effectively by using progressive sequential tree approach. Here we have modified the existing progressive sequential tree to handle the support of an item by adding entry of support for each of the item in addition to the existing field's item-label, sequence-id and time-stamp.

## 2.2 RELATED RESEARCH

### 2.2.1 THE SEQUENTIAL PATTERN MINING

There are many researches about mining sequential patterns in a static database. It was first addressed by Agarwal and Srikant [1]. In general sequential pattern mining algorithms can be classically categorized into three classes. (i) Apriori based horizontal partitioning methods such as Generalized Sequential Pattern mining [4], which adopts multiple-pass candidate generation and test approach in sequential pattern mining. (ii) Apriori based vertical partitioning methods such as Sequential Pattern Discovery using Equivalent classes [5], utilizes combinatorial properties to decompose the original problem into smaller sub-problems that can be independently solved in main memory using efficient lattice search and simple join operations. (iii) Projection based pattern growth algorithms such as prefix-projected sequential pattern mining algorithms [6], which represents the pattern growth methodology and finds the frequent items after scanning database once. In addition to the traditional algorithms there are many which include closed sequential pattern mining [7], maximal sequential pattern mining [8] and constraint sequential pattern mining [9].

### 2.2.2 INCREMENTAL SEQUENTIAL PATTERN MINING

The incremental sequential pattern mining algorithms resolve major drawback of the sequential pattern mining algorithms such as mining the patterns from up-to-date database without deleting the obsolete. The key algorithms of incremental sequential pattern mining are: Parthasarathy et al. [11]**,** developed an incremental mining algorithm ISM by maintaining a sequence lattice of an old database. Sequence lattice includes all the frequent sequences and all the sequences in the negative border. Later Masseglia et al. [10], proposed another incremental algorithm ISE for incremental mining of sequential patterns when new transactions are added to the database. This algorithm adopts candidate generation and test approach. Hang Cheng et al. [3], presented Incspan algorithm which mines sequential pattern over an incremental databases. The limitation of these algorithms is its inability to delete the obsolete data.

### 2.2.3 PROGRESSIVE SEQUENTIAL PATTERN MINING

Progressive sequential pattern mining is a generalized pattern mining methodology that brings out the most recent frequent sequential patterns. This algorithm works both static as well as dynamic changing databases and is unaffected by the presence of obsolete data. The patterns are not affected by the old data. This algorithm uses the sliding window to progressively update sequences in the database and accumulate the frequencies of candidate sequential patterns as time progresses. The sliding window called period of interest determines the time stamp over which the algorithm is currently working.

**Definition 2:** Period of Interest (*POI*) is a sliding window. The length of the *POI* is a user specified time interval. The sequences having elements whose timestamps fall into this period *POI*, contribute to | *DB* | for sequential patterns generated at that timestamp. On the other hand, the sequences having elements with timestamps older than those in the *POI* are pruned away from the sequence database immediately and do not contribute to the | *DB* | thereafter [3].

There are few proposals on progressive sequential pattern mining. The initial study on progressive sequential pattern mining was proposed by Jen W. Huang et al. [3], it gives the complete details of progressive tree concept to capture the dynamic nature of data over a period of interest.

## 3. PROPOSED WORK

To solve the problem of progressive sequential pattern mining having support coupled items, we modify the progressive sequential tree [4], in such a way that it can accommodate supports so we can get patterns with new support. The data structure we use to construct this tree is M-ary tree.

### 3.1 DATA STRUCTURE

M-ary tree structure is used to store the details of progressive database. The nodes of the tree are broadly classified as root node and common node. The root node consists of header, which links with common nodes. Each common node holds the information such as item-name, sequence-id, time stamp and support of an item. Here item -name is the label of an item which is associated with integer number, denotes the support count of an item by default support count is one. Sequence-id stores the list of sequence items to represents the sequence containing this element. Each sequence-id in the sequence list is marked by a corresponding time stamp.

#### 3.1.1 ADDING NEW ITEMS TO M-ARY TREE

At each timestamp the insertion of elements in to the M-ary tree at time t results in an updated tree for time t+1. The algorithm traverses the tree in at time t in a post order. The algorithm continues until there is new data in the progressive database. Whenever a series of elements appear in a sequence, path from the root is created labeled by the respective elements of the pattern with the corresponding sequence number, called the candidate pattern. If a path already exists the concerned fields of the nodes are updated with respective information.

#### 3.1.2 DELETING OBSOLETE ITEMS FROM M-ARY TREE

An obsolete element (i.e., element which lies out of the period of interest) and a node having no sequence number in its sequence list are pruned from the sequential list of the node and the M-ary tree respectively.

### 3.2 MINING FREQUENT PATTERNS FROM PROGRESSIVE DATABASES

The main idea of sequential pattern mining is to utilize the M-ary tree to store all sequences from one period of interest to another. When receiving an item at time stamp say t+1, the algorithm traverses the original M-ary tree of time stamp t in post order to delete the obsolete elements from the updated current sequences in and insert newly arriving elements in a progressive databases i.e, whenever a series of elements appear in a sequence, path from the root is created labeled by the respective elements of the pattern with the corresponding sequence number, called the candidate pattern. If a path already exists then the concerned fields of the nodes are updated with respective information. The time stamp for each node of the candidate sequential pattern is marked according to the starting element of the candidate pattern. An obsolete element is the element which lies out of period of interest and a node having no sequence number in its sequence list are pruned from the sequence list of the node and the M-ary tree as given in algorithm below. Thus we can ensure that there are only up-to-date candidate patterns in M-ary tree [3].

**Procedure traverse (currentTime , PS)**

for(each node of PS in post order)
  if(node is Root)
    for(ele of every seq in eleset)
      for(all combination of elements in the ele)
        if(element == label of one of the node.child)
          create a new sequence with currentTime and support
        else
          create a new child with element, seq, currentTime and support
  else
    for(every seq in the seq_list)
      if(seq.timeStamp <= currentTime - POI)
        delete seq from seq_list and continue to next seq
      if(there is new ele of the seq in eleSet)
        for(combination of elements in the ele)
          if(element is not on the path from Root)
            if(element == label of one of the node.child)
              create a new sequence with seq.timestamp and support
            else
              create a new child with element, seq, timestamp and support
    if(seq_list.size == 0)
      delete this node and its entire child from its parent
  if(seq_list.size >= support*(sequence Number))
    Output the label of the path from Root to node as a SP but if there is an item in between with two support values then take one with higher support value

Figure 1. Proposed Algorithm

### 3.3 THE CONCEPT

The underlying idea of the proposed work is explained with the progressive database given in Fig.2. Where S01, S02…Sn represents different sequence id's. A, B, C and D are the items and $t_1, t_2 \ldots t_n$ represents the timestamps. Every sequence contains a series of elements which are coming in different timestamps and each element contains single or multiple elements associated with support. For instance sequence S01 has elements 2A at timestamp $t_1$. B at timestamp $t_2$. At the bottom of the Fig. 2 $Db^{p,q}$ represents the subset of items from timestamp p to q.

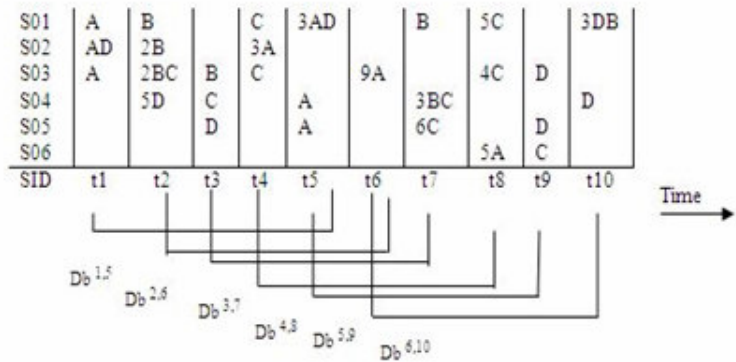

Figure 2. An Example Database

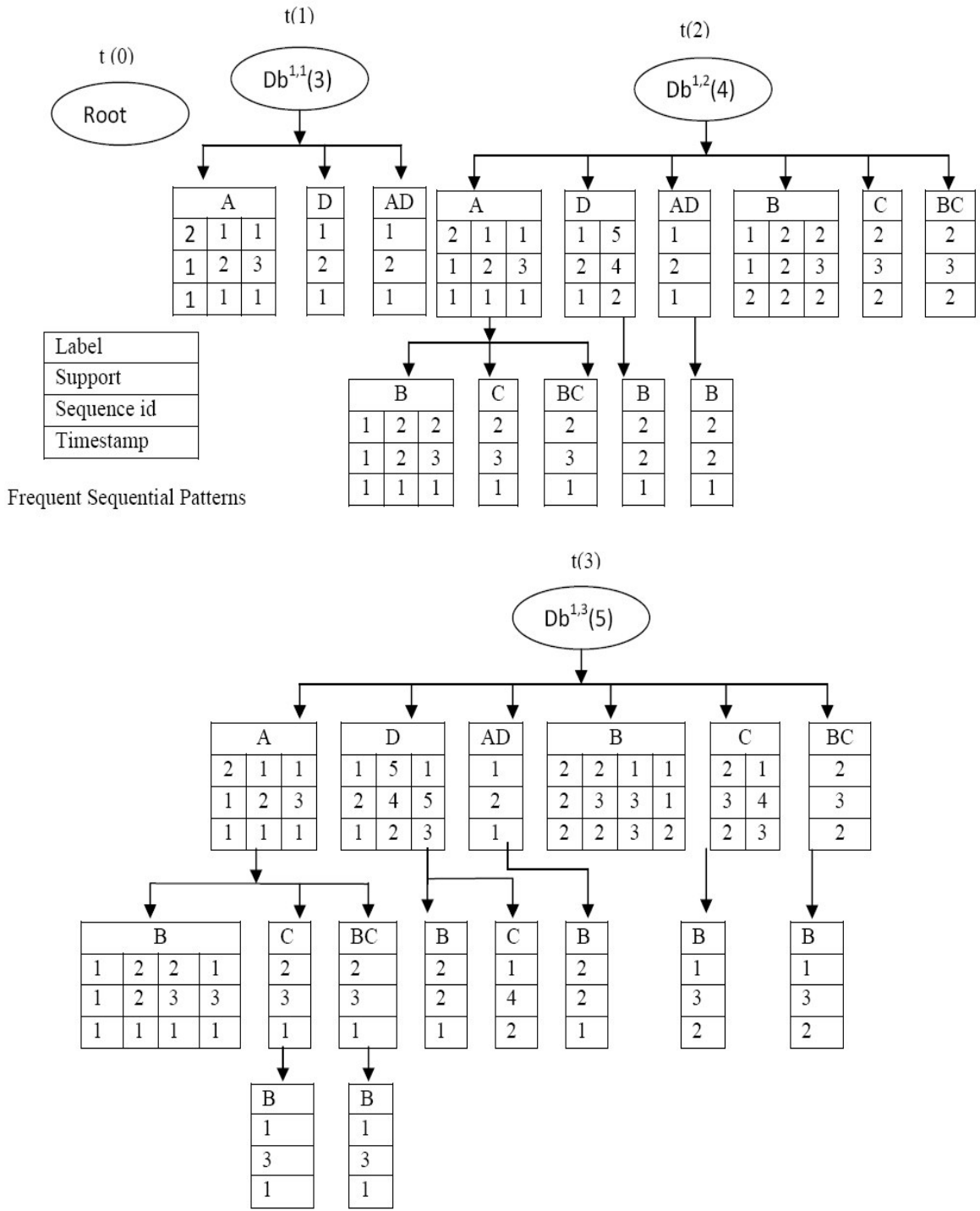

Figure 3. Mary tree of Example Database (t0~t3)

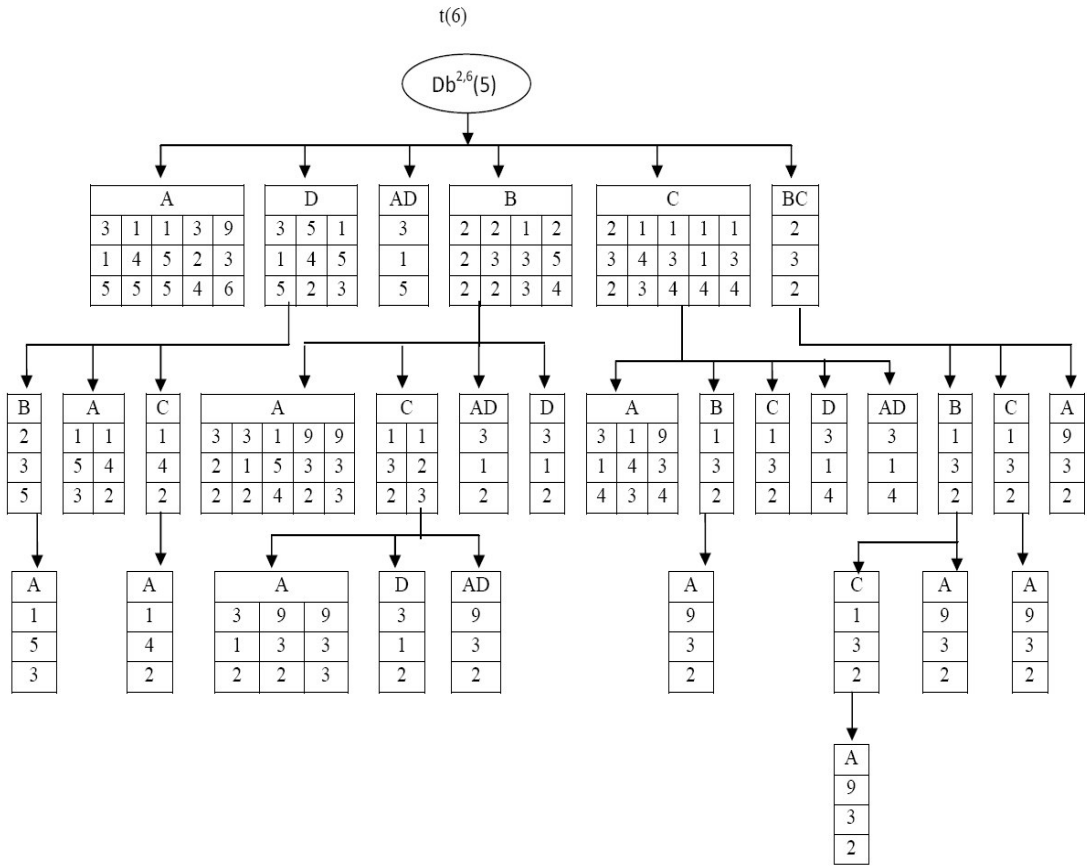

Figure 4. Mary Tree of Example Database ( t2~t6)

## 4. RESULT

Our proposed algorithm is very effectively working for test dataset and we have analysed the test dataset for different parameters like period of interest, execution time etc.

### 4.1 IMPACT OF PERIOD OF INTEREST (POI) ON EXECUTION TIME

Execution time is the time required to execute all the instructions in the proposed algorithm. It can be noted that execution time is directly proportional to POI. The reason which explain this is, the increase in time required to process the expanded data structure which would be required to store the candidate patterns.

### 4.2 IMPACT OF POI AND MINIMUM SUPPORT OVER NUMBER OF PATTERNS

Number of patterns is dependent on both period of interest and the minimum support. As from graph Fig.5, we can see that as the period of interest increases, the number of patterns also increases. This is because as the period of interest increases, the algorithm has more items to process and so they give more number of patterns. Also in case of minimum support Fig. 6, we can observe that as the minimum support value increase, the number of patterns decreases.

The reason behind this is, patterns having lower support will not be frequent as the support value is less than the minimum support value.

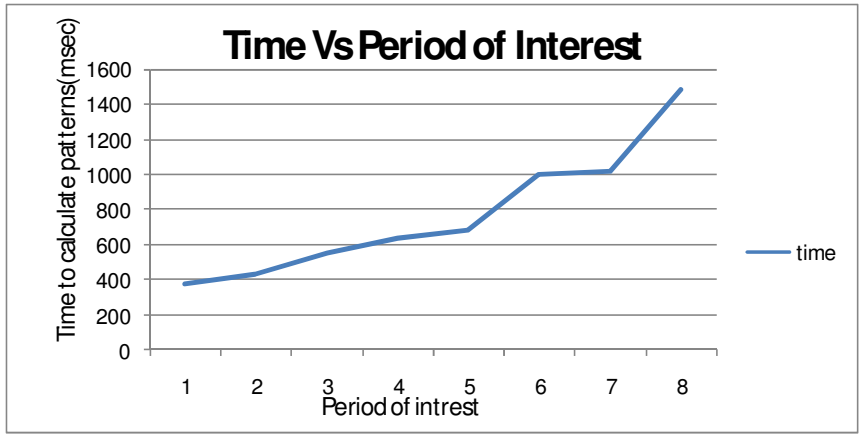

Figure 5. Impact of POI on Execution Time

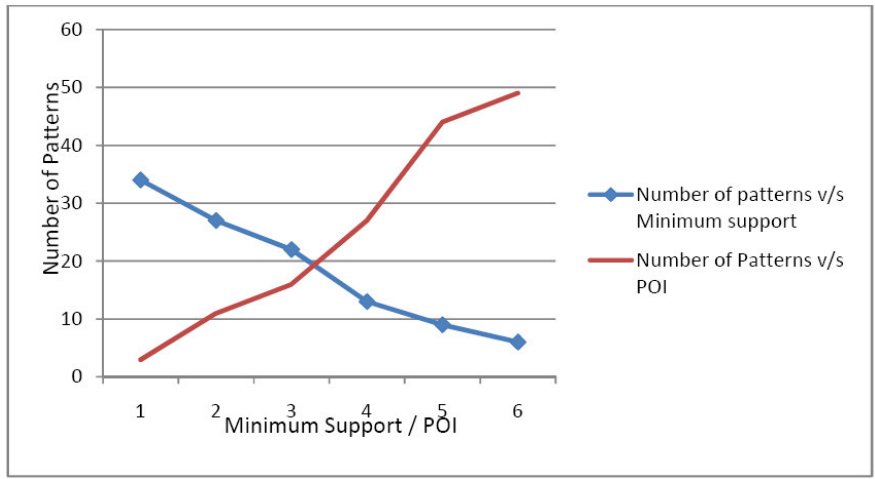

Figure 6. Impact of POI and minimum support over number of patterns

### 4.3 COMPARISON BETWEEN BOOLEAN ITEMS VERSUS SUPPORT COUPLED ITEMS IN PROGRESSIVE DATABASE

In case of progressive database having support coupled items, the frequent patterns must have their support greater than that of the minimum support. Whereas in case of boolean items, only the presence and absence of patterns matters. Thus some of the patterns which were frequent in progressive database having boolean items may not be frequent in support coupled items, as their support count is less than the minimum support. Hence the number of patterns in progressive database having boolean items always greater than or equal to the number of patterns in support coupled items which we can verify from the following graph.

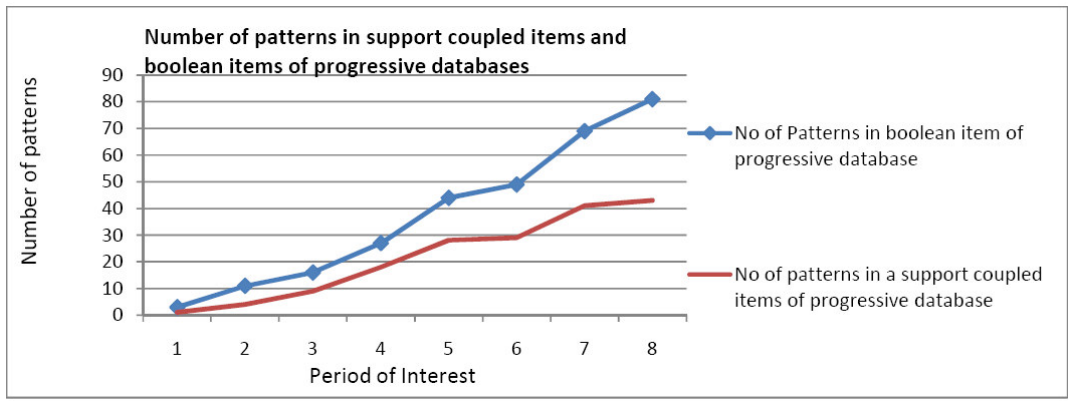

Figure7. Comparison between Boolean items versus support coupled items in progressive databases

## 5 CONCLUSIONS

The proposed novel work mine frequent items associated with support coupled items in progressive sequential databases. The major constraints in mining frequent patterns of progressive sequential databases are that, it should consider the most recent items and they are scanned only once. To achieve this, we have modified progressive sequential tree by using M-ary tree data structure mapping scheme. In this, we record the items over a user defined period of interest, which holds the information such as support coupled item-id, sequence-number and timestamp of each of the items. M-ary tree is very efficient not only in adding new items but also in deleting obsolete elements in only single scan. To remove the obsolete items, we have used apriori's principle to prune the M-ary tree. It is working well with tested dataset. We also successfully analysed the effect of different parameters on the algorithm like period of interest, minimum support etc. Also we compared the performance issue on number of patterns of Boolean versus support coupled item in the progressive sequential databases.

## REFERENCES

[1] R. Agarwal and R. Srikanth, (1991) "Mining Sequential Patterns," In: 11th Int'l Conf. Data Eng. (ICD'95), pp. 3-14.

[2] J. Han and M. kamber, (2000) "Data Mining: Concepts and techniques," Series Editor Morgan Kaufmann Publishers, ISDN 1-55860-489-8.

[3] Jen W. Huang, Chi Y. Tseng, Jian C.Ou and Ming S. Chen,(2008) "A general Model for Sequential Pattern Mining with a Progressive Databases," Int'l. J. Knowledge and Data Eng., vol. 20, No. 9, pp.1153-1167.

[4] R.Srikant and R. Agrawal, (1996) "Mining Sequential Patterns: Generalization and Performance Improvements," In: 5th Int'l. Conf. on Extending Database Technology, pp.3-17, Avignon, France.

[5] M.J. Zaki, (2001) "SPADE: An Efficient Algorithm for Mining Frequent Sequences," J. Machine Learning, vol. 42 No. 1-2, , pp.31-60.

[6]  J. Pei, J. Han, H. Pinto, Q. Chen,U. Dayal and M.C. Hsu, (2001) "PrefixSpan: Mining Sequential Patterns Efficiently by Prefix- Projected pattern Growth," In: 12$^{th}$ Int'l. Conf. on Data Eng., pp. 215-224, Germany.

[7]  S. Cong, J.han and D. Pandu, (2005) "Parallel Mining of Closed sequential Patterns," In: 11$^{th}$ ACM SIGKDD Int'l Conf. Knowledge Discovery and Data Mining (KDD'05), pp. 562-567.

[8]  C. Luo and S.M. Chung, (2005) " Efficient Mining of Maxmal Sequential Patterns using Multiple Samples," In:. 5$^{th}$ SIAM Int'l Conf. Data Mining (SDM).

[9]  Y. Hirate and h. Yamana, (2006) "Sequential Pattern Mining with Time Interval," In: 10$^{th}$ Pacific – Asia Conf. Knowledge Discovery and Data mining (PAKDD'06), pp. 775-779,

[10] F. Masseglia, P. Poncelet and M. Teissere, (2003) "Incremental Mining of Sequential Pattern in Large Database," J. Data Knowl. Eng., 46(1), pp. 97-121.

[11] S. Parthasarathy, M. Zaki, M. Ogihara and S. Dwarakadas, (1999) "Incremental and Interactive Sequence Mining," In: the 8$^{th}$ Int'l. Conf. on Information and Knowledge Management (CIK '99).

[12] H. Cheng, X. Yan and J. Han, (2004) "INCPAN: Incremental mining of Sequential Patterns in Large Database," In: 10$^{th}$ ACM SIGKDD Int'l. Conf. Knowledge Discovery and Data Mining (KDD'04), pp. 527-532.